# Quantum, Diplomacy, and Geopolitics

Strategic imperatives for defence and security in the emerging quantum era

Axel Ferrazzini

December 2025

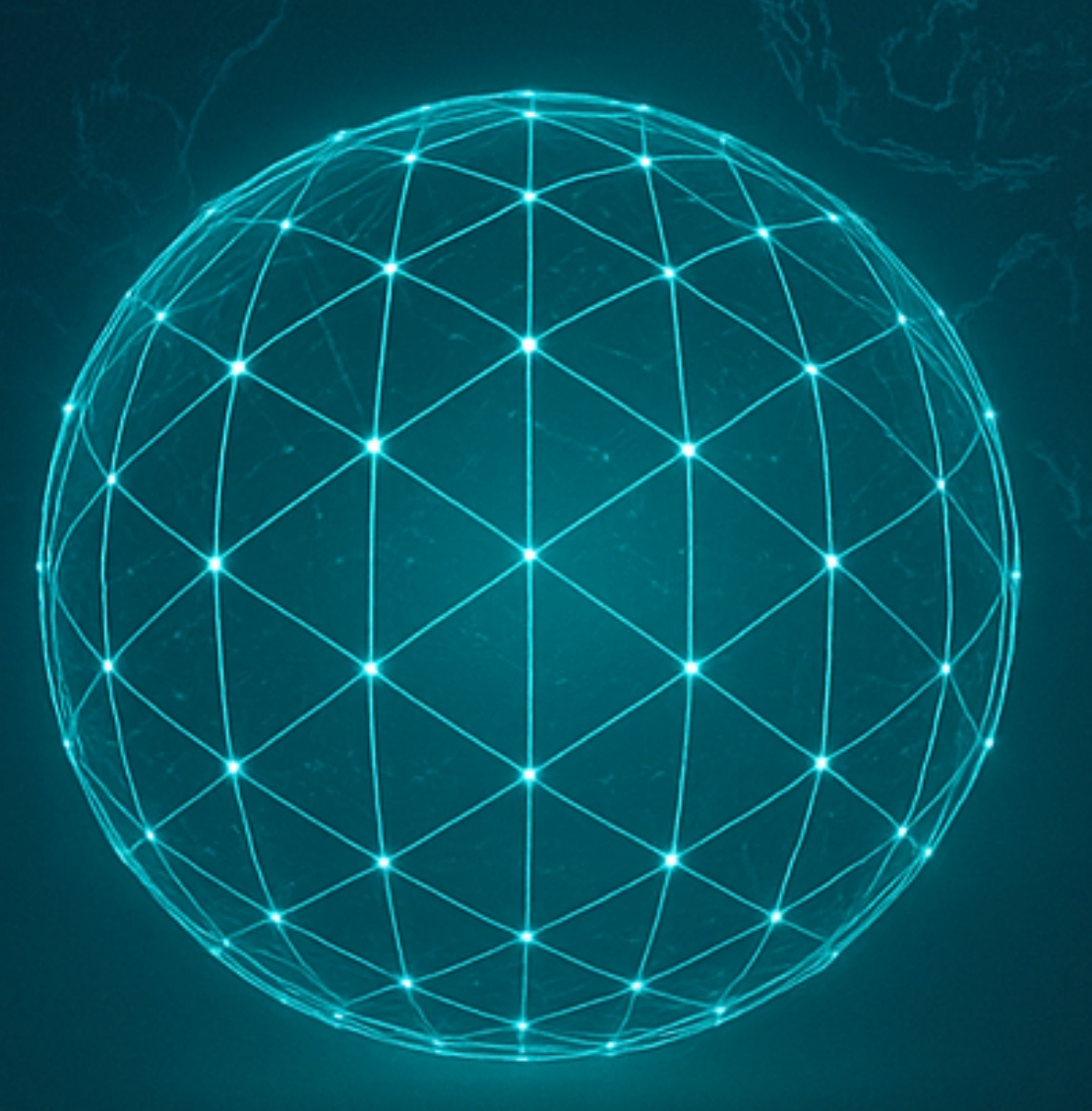

# Quantum, Diplomacy, and Geopolitics

**Strategic imperatives for defence and security in the emerging quantum era**

Axel Ferrazzini[1], December 2025, version 1.6

## Executive Summary

Quantum technologies—spanning communication, sensing, computing, and cryptography—are rapidly emerging as critical paths of geopolitical competition and strategic defence innovation. Unlike traditional technological advances, quantum introduces novel capabilities that fundamentally disrupt established norms of security, intelligence, and diplomatic engagement. This strategic analysis explores the evolving quantum landscape through the dual lenses of diplomacy and geopolitics, with specific implications for defence leaders, policymakers, and industry stakeholders.

The benefits and challenges of quantum technologies are examined from a diplomatic and geopolitical perspective to help leaders make informed strategic decisions.

Leading powers now recognise quantum as a domain where technological leadership directly translates to geopolitical influence, compelling an intense race for dominance alongside new forms of multilateral diplomacy aimed at managing both risks and opportunities. Quantum technologies do not all have the same operational maturity, but technological progress is accelerating. Post-quantum cryptography demands immediate action—every encrypted communication created today may be harvested and decrypted within the decade by adversaries equipped with quantum capabilities.

This analysis presents four core strategic imperatives for Europe:

(1) Accelerate investment in quantum ecosystems while balancing competition with selective cooperation;
(2) Strengthen or establish quantum diplomacy frameworks to manage dual-use technology risks;
(3) Balancing post-quantum cryptography migration and quantum key distribution transition to build a global secure critical infrastructure, including satellites; and,
(4) Integrate quantum capabilities into defence strategy and interoperability standards.

Distinctions are drawn between Europe at large, NATO, the European Union, and the United Kingdom when appropriate.

Forward-thinking organisations that act decisively will secure strategic advantage; those that delay will face exponentially higher costs and unacceptable security vulnerabilities.

[1] Managing Partner at GovStrat, a Brussels-based consulting firm specialised in quantum technology, policy development, standardization and intellectual property strategy. Head of the Quantum major at EPITA, a European leading Computer Science engineering school in Paris. Academic fellow at Bocconi University, Milan. Member of the Board of ETSI. Any comments may be directed to info@govstrat.eu

## Acknowledgements

I would like to thank Olivier Ezratty, Ludovic Perret and Bosco d'Aligny for accepting to review and offer improvements to this strategic analysis.

# 1. Introduction: Quantum as a new domain of strategic competition

For most of the post-Cold War era, technological competition has followed predictable patterns—nations race to miniaturise semiconductors, accelerate data processing, and achieve network dominance. Quantum technologies represent a fundamental departure from this paradigm and do not merely extend existing capabilities; they introduce novel physical principles that render entire categories of security measures obsolete and create unprecedented intelligence and defence opportunities.

The quantum second revolution unfolds across multiple, interconnected domains:

- **Quantum communication** systems promise theoretically unbreakable encryption via quantum key distribution, fundamentally altering intelligence skills and diplomatic security.
- **Quantum sensing** capabilities will enable detection precision previously impossible—from submarine tracking to precision positioning in GPS-denied environments.
- **Quantum computing** promises exponential advances in optimisation, simulation, and overall calculation, with applications spanning materials science, pharmaceutical development, and logistics.
- **Post-quantum cryptography** standards now provide an urgent solution to the "harvest now, decrypt later" threat, which renders today's encrypted data vulnerable to tomorrow's quantum adversaries[2].

What distinguishes quantum's geopolitical significance is its double character: profound scientific collaboration potential coexists with deep strategic competition. The same quantum technologies that scientists worldwide publish and discuss openly also represent potential challenges to national security strategies. This opposition creates challenging diplomatic complexity.

# 2. The quantum landscape: Four pillars of strategic capability

## 2.1. Quantum communication, reshaping intelligence and diplomacy

Quantum key distribution (QKD) leverages quantum-mechanical principles to detect eavesdropping with certainty. Any attempt to intercept quantum states fundamentally alters them, alerting communicating parties to compromise. This capability eliminates the fundamental asymmetry that has defined espionage for centuries: the interceptor's ability to read communications without detection.

China is pioneering the deployment of quantum-secured communications networks to connect government and financial institutions while also collaborating with Russia[3] and BRICS[4] The EU's Quantum Internet Alliance[5] aims to develop quantum network infrastructure across member states, both satellite and ground networks. These deployments signal a strategic transition: quantum communication moves from theoretical advantage to operational reality, and nations that lack it may face intelligence disadvantages[6].

Quantum communications will play a critical role in interconnecting quantum endpoints, such as quantum sensors and quantum computers. By meshing quantum endpoints, quantum communications will increase computing power drastically.

---

[2] https://www.bcg.com/publications/2025/how-quantum-computing-will-upend-cybersecurity

[3] https://quantumzeitgeist.com/china-and-russia-successfully-test-hack-proof-quantum-communication-link-paving-way-for-brics-network/

[4] https://thequantuminsider.com/2025/03/14/china-established-quantum-secure-communication-links-with-south-africa/

[5] https://quantuminternetalliance.org

[6] https://www.bcg.com/capabilities/digital-technology-data/emerging-technologies/quantum-computing

## 2.2.Quantum sensing: Reshaping the military balance

Quantum sensors harness quantum properties to achieve sensing precision orders of magnitude beyond classical systems. Applications will include:

- **Quantum radar and imaging** for enhanced submarine detection and through-foliage surveillance.
- **Atomic clocks, gravimeters** enabling precision navigation independent of GPS, critical in contested electromagnetic environments.
- **Gravimeters and gradiometers** for detecting subsurface structures and concealed military assets.

NATO's quantum strategy explicitly prioritises quantum sensing applications in positioning, navigation, timing, and undersea detection[7]. NATO recognises that quantum-enabled sensors shift tactical and strategic advantage. Adversaries unable to field quantum sensing systems will operate at increasing disadvantage.

## 2.3.Quantum computing: From hype to operational impact

Quantum computing progress follows an exponential trajectory. Today's quantum computers are Noisy Intermediate-Scale Quantum (NISQ) computers, which are devices with tens to a few hundred qubits that are limited by noise, errors, and a lack of full quantum error correction, making them suitable for exploring quantum algorithms and applications with some hybrid classical integration but not yet capable of fault-tolerant computation or large-scale advantages. IBM's roadmap projects scaling from today's hundreds of physical qubit systems to processors exceeding a few hundred logical qubit systems with interconnected processors within five years.[8,9] The difference between a physical and a logical qubit is of the utmost importance.[10] At this scale, quantum computers should not solve optimisation problems beyond the reach of classical supercomputers before 2030. Still, there is no certainty that the development of Fault-Tolerant Quantum Computers (FTQC) will not accelerate faster. Therefore, even if it is not within a decade, quantum computers will solve optimisation problems beyond the reach of classical supercomputers, with applications spanning military logistics, cryptanalysis, weapons design, and strategic simulation in the near future.

**Fault-Tolerant Quantum Computers: Unleashing the full power of quantum computing**

Building fault-tolerant quantum computers is a high-stakes, decade-scale bet because it is the only credible path to systems that can consistently outperform classical supercomputers on problems that matter in many domains at an industrial scale. Today's noisy devices are powerful for learning and ecosystem building. Still, they are fundamentally prototypes: without full error correction, they remain fragile, hard to scale, and very limited in the size and depth of algorithms they can run. Moving to FTQC means mastering everything from qubit physics and cryogenics to control electronics, energy efficiency, and software stacks.[11] It also requires navigating a complex global vendor landscape, with multiple competing qubit technologies and large incumbent and startup players all racing to demonstrate credible roadmaps to logical qubits.

The strategic stakes are equally significant at the ecosystem and geopolitical level: whoever cracks FTQC first will reshape competitive advantage in drug discovery, advanced materials, optimisation-heavy industries, and possibly parts of cybersecurity, reinforcing national and regional technology sovereignty. For governments, that translates into questions of research funding, standards, export controls, and industrial policy; for corporates, it is about timing: investing early enough to build capabilities and data assets. Given the scale and cost, FTQC will not be a “winner-takes-all” market but a stack play in which value pools across enabling technologies, software, cloud delivery, and applications emerge.

[7] https://www.nato.int/en/news-and-events/articles/news/2024/01/17/nato-releases-first-ever-quantum-strategy

[8] https://www.ibm.com/roadmaps/quantum/

[9] https://www.oezratty.net/Files/Conferences/Olivier%20Ezratty%20ARTEQ%20Quantum%20Computing%20Roadmaps%20Nov2025.pdf

[10] https://en.wikipedia.org/wiki/Physical_and_logical_qubits & https://www.quera.com/glossary/logical-qubit

[11] https://www.oezratty.net/wordpress/2025/understanding-quantum-technologies-2025/

The practical implication for senior leaders is clear: treat FTQC as a strategic option with structured bets, portfolio hedging across technologies, and a strong focus on talent and partnerships, rather than as a short-term IT investment.

Unlike quantum communication and sensing—where partial, near-term capabilities offer immediate advantage—quantum computing remains primarily in the research phase despite the hype. However, this reality has not dampened competition. The United States, China, and the European Union have each committed multi-billion-euro quantum computing development programs. (still much lower than their respective investments in artificial intelligence) The strategic calculus is clear: first-mover advantage in quantum computing can lead to long-term technological dominance.

### 2.4.Post-quantum cryptography: The immediate imperative

While quantum computing remains nascent, the cryptographic threat it poses is immediate and acute. The "harvest now, decrypt later" scenario represents perhaps the most pressing quantum-related security vulnerability affecting governments and industry today. Adversaries currently harvest and store encrypted communications—classified intelligence, business secrets, weapons designs—with confidence that decryption remains impossible. However, once quantum computers achieve sufficient capability—estimated by leading analysts around 2035—these harvested communications become readable[12].

The mathematics underlying this threat is inevitable. Quantum computers will excel at integer factorisation[13] and discrete logarithm problems—the exact mathematical foundations underlying RSA and elliptic curve cryptography that protect most sensitive communications today. Classical computers cannot solve these problems at scale, but quantum computers will.

Recognising this threat, the US National Institute of Standards and Technology (NIST) initiated a standardisation competition in 2016. A total of five approved post-quantum cryptographic algorithms were announced: In 2022, CRYSTALS-Dilithium (renamed ML-DSA), FALCON (renamed FN-DSA), SPHINCS+ (renamed SLH-DSA), CRYSTALS-Kyber (renamed ML-KEM), and then in 2025, HQC (whose official name has not been chosen yet). These algorithms derive security from mathematical problems assumed intractable for both classical and quantum computers—providing the foundation for quantum-resistant security[14].

## 3. Quantum diplomacy: Navigating cooperation and competition

Quantum technologies present a unique diplomatic challenge: they simultaneously demand international cooperation and enable strategic competition. The challenge for diplomacy is managing this paradox without sacrificing national security or innovation.

[12] https://www.bcg.com/publications/2025/how-quantum-computing-will-upend-cybersecurity

[13] https://arxiv.org/abs/2505.15917 & https://arxiv.org/pdf/2507.12511

[14] https://www.nist.gov/news-events/news/2024/08/nist-releases-first-3-finalized-post-quantum-encryption-standards

## 3.1.The Fragmentation Risk

The quantum era risks creating fragmented, incompatible ecosystems due to its dual-use nature. Chinese quantum communication networks may not interoperate with European systems. American quantum computing standards may diverge from EU approaches. This fragmentation carries profound implications:

- **Intelligence and allied cooperation** depend on secure, interoperable communications. Incompatible quantum systems would degrade NATO interoperability and intelligence sharing.
- **Multinational supply chains** require compatible standards. Fragmentation increases costs and delays the deployment of critical infrastructure. (see below explanations regarding export control).
- **Scientific progress** depends on open publication and collaboration. Strategic classification of quantum research may limit scientific advancement, benefiting all nations.

**Export control regulations and agreements: Risks of fragmentation**

ITAR (International Traffic in Arms Regulations) is a U.S. regulatory regime that controls the export and import of defence-related technologies, including quantum technologies with potential military applications, such as quantum encryption and quantum computing, to prevent sensitive technologies from falling into the hands of adversaries and thereby impacting research collaboration and international commercialisation.

The Wassenaar Arrangement is a multilateral export control agreement among 42 countries that sets guidelines for controlling the export of dual-use goods and technologies, including certain quantum technologies, to promote transparency and prevent destabilising transfers that could be used in military or intelligence contexts.

Both ITAR and the Wassenaar Arrangement impose licensing and restrictions on the transfer of advanced quantum technologies, which can slow down international cooperation, complicate supply chains, and require companies and research institutions in the quantum ecosystem to navigate complex compliance regimes. These controls, while addressing security concerns, also risk limiting access to critical quantum components, software, and know-how, and may shape national strategies around talent, investment, and industry development in quantum computing, sensing, and communications technologies.

**Concentration of excessive power in a single country: Further risks of fragmentation**

If the United States (or China) concentrates excessive power in quantum technologies, the UK, the EU, and Europe at large could face significant strategic and economic vulnerabilities. This concentration risks creating a technological dependency that undermines Europe's autonomy in critical domains such as secure communications, defence capabilities, and advanced computing. Given the strong relationship among the United Kingdom, the European Union, and the United States in quantum technologies, in the hypothetical situation where the United States decides to hinder this relationship, the United States' dominance might lead to restricted access to cutting-edge quantum technologies and prevent European entities from accessing enabling technologies, raw materials and advanced capabilities. Such dependencies could impair Europe's ability to safeguard sensitive data, optimise industrial and military applications, and catalyse innovation within its borders, potentially widening a strategic technology gap and compromising long-term sovereignty.

Moreover, this imbalance could exacerbate geopolitical tensions, forcing Europe into complex alignments or alliances that might not entirely reflect its strategic interests. The United States' dominance in quantum technologies may also enable enhanced intelligence and cyber advantages, weakening European cybersecurity resilience. Economic consequences may include loss of market opportunities, brain drain, and limited growth for quantum startups and industries in Europe. Ultimately, it could also make FTQC systems *made in Europe* too expensive or even impossible. To counteract these effects, Europe would need to intensify investment in a collaborative, integrated quantum ecosystem focused on strategic sovereignty, strengthen digital independence frameworks, and actively engage in international quantum diplomacy to shape a more balanced, multipolar quantum technology landscape.

## 3.2.Diplomatic Frameworks Emerging

Several multilateral initiatives address quantum diplomacy:

**NATO's Quantum Strategy** (approved January 2024) represents the first alliance-level commitment to quantum capabilities, emphasising transatlantic cooperation, interoperability standards, and defence preparation[15]. The strategy commits NATO to establishing a Transatlantic Quantum Community engaging government, industry, and academia.

By pooling resources and capabilities in quantum research and development, NATO also helps its European members, including the UK and the EU, reduce their dependence on external quantum technology providers, such as the US or China, thereby enhancing European digital sovereignty within the alliance framework. NATO actively supports the integration of quantum-resistant cryptography across its infrastructure to defend against future quantum-enabled cyber threats, thereby strengthening collective cybersecurity. Moreover, NATO-backed initiatives encourage responsible innovation that reflects democratic values, maintain technological superiority while reinforcing strategic trust between allies, and offer a platform for Europe to influence global quantum developments. NATO's strategic coordination tries to mitigate the risks of technological fragmentation and geopolitical rivalry, securing a multipolar balance in quantum technology leadership across the Atlantic.

**EU Strategic Autonomy** initiatives frame quantum as essential to European sovereignty. The Quantum Flagship, the Quantum Europe Strategy[16] and the expected 'Quantum Act' for 2026, emphasise European technological independence rather than reliance on non-EU suppliers.

**United Nations and UNODA** discussions explore quantum's implications for arms control and strategic stability, though formal agreements remain nascent[17].

**The United Kingdom: Leading diplomatic framework development.**

The UK has made substantial efforts to establish a quantum diplomacy framework, linking scientific leadership and security priorities with international engagement: The UK's National Quantum Strategy (2023) explicitly integrates international collaboration as a core pillar. It targets secure communications, standards setting, and the rapid deployment of quantum technologies with trusted allies and partners.[18]

In November 2025, the UK signed a new Technology Partnership with the United States focused on quantum technologies (alongside nuclear and AI). The partnership features joint research, shared industrial exchanges, benchmarking, AI-quantum convergence, and collaborative early-deployment programs, aiming to be a cornerstone in transatlantic quantum cooperation.[19]

The UK has formalised research and standards arrangements (MOUs) with five quantum-leading countries: Denmark, the Netherlands, the United States, Canada, and Japan. It also co-chairs the NMI-Q (National

---

[15] https://www.nato.int/en/news-and-events/articles/news/2024/01/17/nato-releases-first-ever-quantum-strategy

[16] https://digital-strategy.ec.europa.eu/en/library/quantum-europe-strategy

[17] https://open-quantum-institute.cern/diplomats-engage-in-quantum-diplomacy-game-at-the-un-office-for-disarmament-affairs-unoda/

[18] https://www.gov.uk/government/publications/national-quantum-strategy/national-quantum-strategy-accessible-webpage

[19] https://www.quantumworldcongress.com/news-and-updates/united-kingdom-unveils-next-phase-of-quantum-strategy-with-500m-mission-funding-and-new-ukus-tech-partnership

Metrology Institute – Quantum) international quantum standards initiative with the US, involving G7 nations and Australia.[20]

The UK participates actively in global quantum policy and governance forums, including the Quantum Development Group and other efforts to coordinate quantum technology standards, responsible innovation, and infrastructure deployment worldwide.[21]

The NCSC launched a pilot consultancy scheme and partnerships for post-quantum cryptography migration, targeting both supply chain resilience and the creation of quantum-safe digital infrastructure for public and private sectors.[22]

The UK's approach combines proactive leadership in quantum research, international coalition-building, and regulatory adaptation to ensure both competitive advantage and digital sovereignty in alignment with broader diplomatic, economic, and security interests.

**G7: A common vision for the future of quantum technologies.**

The Kananaskis Common Vision for the Future of Quantum Technologies, endorsed by G7 Leaders in June 2025, proposes a collaborative diplomatic framework centred on international cooperation among like-minded governments, researchers, industry, and stakeholders to harness quantum technologies' potential in computing, sensing, and communications while mitigating risks to security and data protection. It emphasises joint commitments such as mobilising public-private investments, securing supply chains, fostering workforce development (including STEM education and inclusion of underrepresented groups), promoting trusted ecosystems through interoperability and IP protection, and advancing policy dialogues via a dedicated G7 Joint Working Group on Quantum Technologies to assess societal impacts and enable voluntary joint projects. This non-regulatory approach prioritises open dialogues, best-practice exchanges, and alignment with democratic values over premature global rules, aiming to accelerate commercialisation, innovation, and quantum-resilient measures in critical sectors such as defence and infrastructure.[23]

## 3.3. Standards as diplomatic instruments

International standards development bodies—such as ITU (International Telecommunication Union) and ISO—may be perceived as de facto diplomatic forums where nations negotiate technological compatibility and competitive boundaries. Geopolitical tensions may threaten the creation of international standards. ITU and IEC/ISO could implicitly influence which nations' technological approaches are adopted globally and which are marginalised. This dynamic might turn technical standardisation into a form of geopolitical rivalry. Meanwhile, ETSI (European Telecommunications Standards Institute), an international standards development organisation and European standards organisation, is a neutral entity for defining standards and enabling interoperability, as evidenced by its involvement in establishing GSM and forming 3GPP for the development of 2G, 3G, 4G, 5G, and now 6G.

**De facto standards are also emerging: Risks of vendor lock-in.**

Formal standards development organisations specify open standards to enable interoperability, while companies create de facto standards: they gain popularity and become dominant. Here are two examples of de facto standards:

[20] https://www.gov.uk/government/news/government-support-to-get-quantum-to-work-faster-boosting-uks-health-defence-energy-and-more

[21] https://uknqt.ukri.org

[22] https://www.ncsc.gov.uk/collection/ncsc-annual-review-2025/chapter-03-keeping-pace-with-evolving-technology/migrating-to-post-quantum-cryptography

[23] https://g7.canada.ca/en/news-and-media/news/kananaskis-common-vision-for-the-future-of-quantum-technologies/

IBM's Qiskit is becoming the dominant framework for quantum software developers. It potentially poses several risks despite its convenience and widespread adoption. Primarily, reliance on a single proprietary ecosystem can lead to vendor lock-in, limiting interoperability and restricting innovation by tying developers and organisations to specific hardware architectures and software tools. This concentration can stifle competition and slow the diversification of quantum software approaches necessary to adapt to evolving quantum hardware paradigms. Furthermore, dominance by a single player may create geopolitical vulnerabilities if access, updates, or compatibility are influenced by corporate or national interests that do not align with all global stakeholders. For Europe, this could increase dependency risks, complicating efforts to build open, secure, and resilient quantum ecosystems tailored to its strategic and regulatory requirements.

Nvidia's NVLink is a critical high-speed interconnect for Nvidia GPUs. NVLink is a proprietary hardware communication protocol primarily used to link Nvidia GPUs and other accelerators, enhancing performance for AI and high-performance computing workloads. The potential concern with NVLink centres around ecosystem lock-in to Nvidia's specific hardware and software stack rather than software development frameworks broadly. Organisations heavily reliant on NVLink may face challenges in migrating to alternative hardware architectures or integrating heterogeneous systems, creating a form of vendor lock-in at the hardware level.

## 3.4.Patents as crucial diplomatic instruments

Patents are crucial diplomatic instruments in quantum diplomacy because they not only protect intellectual property but also serve as strategic indicators of technological leadership and geopolitical influence. The European Patent Office (EPO) and OECD "Mapping the global quantum ecosystem" report[24] reveals a surge in quantum computing patent filings worldwide, highlighting intensified global competition and cooperation. Patent families—covering inventions filed in multiple jurisdictions—reflect cross-border collaborations and signal where major players place their bets on quantum technologies. The geographical distribution of patent applicants shows strong regional clusters such as the US, Europe, and Asia, but also reveals increasing efforts to establish dominance via multi-national patent protection. Such patent activity is vital for countries to secure their research investments, control market access, and steer the global quantum innovation ecosystem in line with their national interests and diplomatic strategies. Patents thereby become both economic assets and levers of influence in setting technological standards, negotiating partnerships, and shaping global governance frameworks in emerging quantum domains.[25, 26]

The EPO further underscores that quantum computing, especially in areas such as quantum error correction and AI-driven quantum applications, is a high-growth patent field that involves significant cooperation between companies and academic institutions within and across regions. This collaborative patenting activity represents a form of technological diplomacy where alliances are forged and technological ecosystems are co-developed, reinforcing mutual dependencies and soft power. Governments and industry use patent portfolios strategically to signal capability, attract investment, and enhance bargaining power in international quantum forums. Patents thus function as a bridge between innovation and policy, enabling countries to safeguard sovereign capabilities while engaging in diplomatic efforts to manage risks, foster trust, and collaborate on standards and security in a quantum future.

[24] https://www.oecd.org/content/dam/oecd/en/publications/reports/2025/12/mapping-the-global-quantum-ecosystem_47891dd2/20251217-0001.pdf

[25] https://www.epo.org/en/news-events/events/scaling-quantum-innovation

[26] https://link.epo.org/web/epo_patent_insight_report-quantum_computing_en.pdf

# 4. Geopolitical dynamics: National strategies and strategic competition

## 4.1.The United States: Defence leadership and strategic continuity

The US approach emphasises maintaining technological superiority across all quantum domains. The National Quantum Initiative (launched in 2018)[27] coordinates federal investment, industry partnerships, and academic research. The Department of Defence prioritises quantum sensing and computing for military advantage.

The Defense Advanced Research Projects Agency (DARPA) plays a pivotal role in the US quantum technology procurement landscape through its Quantum Benchmarking Initiative (QBI). Launched in 2024, QBI is a multi-stage program designed to rigorously evaluate and validate quantum computing architectures, aiming to achieve a utility-scale, FTQC by 2033. DARPA invests a significant amount of money per company and fosters collaboration with national labs, industry, and academia to maintain US leadership in quantum technologies by selectively procuring viable quantum systems.

The Quantum Benchmarking Initiative (QBI) is part of a broader procurement and innovation framework where DARPA emphasizes rapid transition from laboratory research to deployable quantum technology. In this context, QBI works closely with initiatives like the Quantum Frontier Project and collaborates with states and other federal agencies to leverage testing infrastructure and expertise. The procurement mechanism focuses on identifying and validating architectures that can deliver strategic advantages, such as enhanced computation for defence applications or secure communication.

Strategically, the US frames quantum as a domain where American technological innovation and transatlantic partnership ensure Western strategic advantage. The US approach combines offensive innovation (racing to quantum capabilities), defensive measures (accelerating PQC adoption), and diplomatic leverage (setting standards through NIST).

The US has established economic dominance through a mix of significant investments, driven by its leading global technology companies, leading venture capital firms, and desirable stock markets. This mix allowed the investment of billions in European startups. Private investment exceeds public investment only in two countries: the US and Denmark.

## 4.2.China: Ecosystemic ambition and strategic self-sufficiency

Despite the lack of transparency, China's quantum strategy emphasises rapid capability development and technological self-sufficiency. China's quantum communication network is said to be more advanced than Western deployments. China is now working on its own PQC standardisation as well.

China frames quantum as central to technological sovereignty. Rather than seeking compatibility with Western systems, the Chinese strategy is said to emphasise developing quantum capabilities independently and offering quantum technologies to Belt-and-Road Initiative participants—creating a potential alternative quantum ecosystem.

China is also investing heavily in quantum sensing, mainly for military applications, such as quantum radars and submarine sensors.

---

[27] https://www.quantum.gov

## 4.3.The UK: First-mover advantage

The National Quantum Computing Centre (NQCC) in the UK plays a central role in the government's procurement and development strategy for quantum technologies. As the national laboratory focused on accelerating quantum computing, the NQCC is jointly funded by UK Research and Innovation (UKRI), with long-term backing secured through a 10-year funding settlement. This funding enables the NQCC to establish and operate world-class quantum computing infrastructure, including hosting a fleet of quantum computers developed by both private companies and in-house scientists. By doing so, the NQCC provides essential testbeds and facilities that lower market-entry risks for developers and accelerate the commercialisation and practical adoption of quantum computing across key sectors, from healthcare to energy and national defence.[28, 29, 30]

In terms of procurement, the NQCC strategically engages with private quantum technology vendors through flexible, tailored contracts. This approach allows the centre to procure and integrate quantum hardware and software that best meet the evolving needs of UK industry and government. The NQCC's collaborations span academia, industry, and government, supporting innovation ecosystems and scaling early-stage quantum companies by providing them access to national-level infrastructure and expertise.

## 4.4.The EU: Strategic sovereignty through coordinated infrastructure and regulation

The EU's quantum strategy[31] reflects a distinctive geopolitical posture: rather than pursuing technological dominance comparable to the US or China, the European Commission emphasises strategic autonomy through coordinated infrastructure development, regulatory harmonisation, and distributed innovation across member states. This approach acknowledges a critical reality—no single European nation possesses the technological and financial capacity to match US or Chinese quantum capabilities—while asserting collective European agency in shaping global quantum standards and governance.

**Quantum computing is the top priority**

Quantum computing is the EU's top priority in its quantum strategy. It is supported by flagship projects such as the Quantum Computers for Datacentres (QCDC) initiative[32], which should deliver cloud-based access to world-class European quantum devices, enabling local innovation while reducing dependency on non-EU providers. The EU's Quantum Europe Research and Innovation Initiative reinforces foundational research, industrialisation, and ecosystem development, alongside the EuroHPC Joint Undertaking, integrating hybrid quantum-supercomputing platforms across member states.

**Quantum communication infrastructure as a strategic priority**

The European Quantum Communication Infrastructure (EuroQCI) represents the cornerstone of EU quantum sovereignty. Unlike point-to-point quantum key distribution systems deployed in China and the US, EuroQCI envisions a distributed, pan-European quantum network infrastructure connecting government, defence, and critical infrastructure institutions across all EU member states. This infrastructure-first approach prioritises interoperability and collective security over the development of individual national capabilities.

**Regulatory harmonisation as a compliance mechanism**

---

[28] https://www.nqcc.ac.uk/updates/uks-industrial-strategy-2025-published/

[29] https://www.ukri.org/news/670-million-boost-gives-certainty-to-uk-quantum-computing/

[30] https://www.nqcc.ac.uk

[31] https://digital-strategy.ec.europa.eu/en/library/quantum-europe-strategy

[32] https://qt.eu/news/2025/2025-08-28_eu-gives-unprecedented-access-to-quantum-computers

The EU tries to transform post-quantum cryptography adoption from a technical recommendation to a regulatory obligation through multiple complementary mechanisms:

**NIS2 Directive (Network and Information Security Directive 2)** [33] establishes mandatory cybersecurity requirements for operators of essential services (energy, transportation, healthcare, finance) and digital service providers. NIS2 explicitly requires a transition to quantum-safe cryptography, with implementation timelines beginning in 2025. This regulatory instrument converts PQC migration from a discretionary business decision to a compliance imperative, creating deterministic market demand for European PQC solutions.

**DORA (Digital Operational Resilience Act)** [34] extends quantum-safe requirements to financial institutions, ensuring the EU's critical financial infrastructure achieves quantum-resistant security. DORA's comprehensive operational resilience framework embeds PQC transition as a non-negotiable requirement for infrastructure modernisation.

**Cybersecurity Act** [35] **and Digital Services Act** [36] complement this regulatory framework, establishing harmonised security standards across digital infrastructure. Collectively, these regulations create a comprehensive compliance ecosystem that imposes the adoption of quantum-safe cryptography across EU institutions, critical infrastructure, financial services, and digital platforms.

**Market and industrial strategy**

An EU regulatory harmonisation creates market advantages for European quantum technology suppliers positioned to deliver compliance solutions. The EU approach leverages regulatory power to create coordinated, standardised markets where European suppliers compete on standardised solutions rather than proprietary advantage.

The Quantum Flagship supports European quantum research across communication, computing, and sensing domains while maintaining emphasis on practical, deployable solutions aligned with regulatory requirements. This research strategy prioritises near-term quantum communication capabilities and cryptographic transition over longer-term quantum computing ambitions.

**Strategic constraints and opportunities**

The EU's approach reflects an honest assessment of geopolitical constraints. Europe lacks the technological concentration of the US and the state capacity for sustained mega-projects characterising China's quantum development. Rather than competing on these dimensions, the EU leverages its genuine strengths: regulatory harmonisation authority, distributed scientific excellence, and multinational coordination mechanisms.

This strategy creates specific opportunities and vulnerabilities. The opportunity lies in establishing global standards—if EuroQCI deliverables become standards and achieve widespread international adoption, European technological approaches will gain global influence. The vulnerability lies in regulatory compliance imposing costs on the EU industry without corresponding technological leadership benefits. If quantum capabilities concentrate in non-EU locations, EU regulatory harmonisation becomes a compliance tax rather than a competitive advantage.

**EU’s differentiations and challenges**

---

[33] https://digital-strategy.ec.europa.eu/en/policies/nis2-directive

[34] https://www.eiopa.europa.eu/digital-operational-resilience-act-dora_en

[35] https://digital-strategy.ec.europa.eu/en/policies/cybersecurity-act

[36] https://digital-strategy.ec.europa.eu/en/policies/digital-services-act-package

The existence of diverse European and national economic incentives, financial benefits, procurement projects and economic assistance creates challenges for startups. Navigating all these administrative hurdles requires resources and time. Some companies must balance their resources between scientific/engineering research and administrative filings. Proposals to simplify the administrative burden in the EU exist, one proposal is the 28th Regime.[37] The EU offers significant research collaborative projects and procurement mechanisms, such as EuroHPC, EuroQCI and QIA, that help nurture the European quantum ecosystem and embed European values[38]. A key opportunity is the opportunity to create an energetic advantage for European quantum computer vendors and shape the global landscape.[39, 40]

**EU Member States also have their own quantum strategic initiatives – here are the most significant**

**France** – The PROQCIMA program[41] is a flagship element of France’s national quantum strategy. Its primary goal is to develop and industrialise two prototypes of universal FTQC with 128 logical qubits by 2032, scaling up to 2048 logical qubits by 2035. It aims at ensuring France’s technological sovereignty by fostering a robust domestic quantum ecosystem, supporting talent attraction, and positioning the country as a world leader in quantum computing innovation and industrial-scale deployment.

**Germany** – The German Aerospace Centre (DLR) plays a crucial role in Germany’s quantum strategy by serving as a key research and development hub that bridges fundamental quantum science with applied technological innovation. DLR’s involvement focuses on advancing quantum communication, quantum sensing, and quantum computing technologies that underpin Germany’s objectives for strategic sovereignty and robust digital infrastructure. It supports the development of quantum-safe encryption methods, satellite-based quantum communication links, and quantum sensors with applications in aerospace, defence, and industrial sectors. This aligns with Germany’s broader national quantum strategy to become a global leader in secure communication and precision sensing, while fostering competitive innovation ecosystems within Europe.[42]

**The Netherlands** – Quantum Delta NL plays an essential role in the Netherlands’ ecosystem-building programme. Quantum Delta NL channels renewed National Growth Fund support into three “catalyst” domains—quantum computing and simulation, a national quantum network, and quantum sensing—underpinned by action lines on research and innovation, ecosystem development, human capital, and societal impact, with a recent pivot from a flagship “House of Quantum” building toward shared cleanroom facilities and pilot-line infrastructure across hubs such as Delft, Leiden, Twente, Amsterdam and Eindhoven to accelerate commercialization and industrial uptake.[43] In parallel, the central government runs a Quantum-Secure Cryptography programme and pilots an operational quantum network using quantum key distribution and post-quantum cryptography to protect ministries’ communications, explicitly framing quantum as both an economic opportunity and a cybersecurity risk and positioning the Netherlands to implement the European Commission’s roadmap for a quantum-safe digital future.

## 4.5. Switzerland: Neutral quantum lead

Switzerland’s national strategy leverages its central European position as a bridge between Western alliances and multipolar powers to safeguard sovereignty through enhanced security investments, cyber resilience, and deepened ties with EU partners and NATO in non-combat domains, while pursuing technological primacy amid

37 https://www.euroquic.org/the-28th-regime-and-innovative-quantum-companies/

38 https://european-union.europa.eu/principles-countries-history/principles-and-values/aims-and-values_en

39 https://quantum-energy-initiative.org

40 https://www.oezratty.net/Files/Conferences/Olivier%20Ezratty%20Q2B%20SV%20FTQC%20Energetics%20Dec2025.pdf

41 https://quantique.france2030.gouv.fr/acces-aux-marches/programme-proqcima/

42 https://www.kas.de/en/single-title/-/content/quantum-technology-and-germany-s-security-policy-a-geopolitical-necessity

43 https://quantumdelta.nl/news/quantum-delta-nl-expands-strategy-with-renewed-national-growth-fund-support

US-China rivalry. This posture is amplified through the CERN-hosted Open Quantum Institute (OQI)[44], a multilateral science diplomacy platform incubated by GESDA[45] and backed by UBS, which democratises global access to quantum computing for societal challenges such as climate modelling and healthcare, fostering inclusive applications through cloud-based public-private quantum resources. OQI leadership integrates CERN's Quantum Technology Initiative coordination, led by experts, alongside industry giants, Swiss powerhouses ETH Zurich and EPFL, and governance from QuIC and EuroQCI affiliates, positioning Switzerland as a neutral vanguard in quantum sovereignty and ethical innovation.

# 5. Defence implications: Quantum-ready military systems

The maturity of quantum technologies has been accelerating over the last decade and will lead to quantum-ready systems despite uncertain timings.[46] Military applications will require proactive policymaking[47], as introduced in this section.

## 5.1.Quantum sensing in military operations

Quantum sensors fundamentally alter military capabilities in multiple dimensions, but some are closer to reality than others:

- **Submarine detection:** Quantum gradiometers and other quantum sensors could detect submarine magnetic signatures and mass anomalies with unprecedented sensitivity, potentially negating stealth advantages that have defined submarine dominance for decades.
- **Precision positioning:** Quantum atomic clocks enable navigation accuracy independent of GPS (also called Quantum non-GNSS navigation), essential in contested electromagnetic environments where adversaries employ GPS jamming[48]. PNT sensing, which refers to the collection and processing of signals for Positioning, Navigation, and Timing information from quantum sensors, is advancing rapidly.
- **Persistent surveillance:** Quantum imaging penetrates concealment—detecting camouflaged forces, hidden installations, and below-surface military assets with previously unattainable precision.
- **Broadband electromagnetic spectrum analysis** involves capturing and processing signals across a wide range of frequencies and is simultaneously useful for intelligence and Electronic Intelligence (ELINT).

NATO's explicit inclusion of quantum sensing in its strategy reflects recognition that quantum-enabled sensing confers decisive tactical and operational advantages[49]. Allies lacking quantum sensing capabilities will operate at escalating disadvantage.

## 5.2.Quantum computing in military applications

Quantum computing enables military applications across multiple domains, most of which are quite long-term:

- **Logistics optimisation:** Solving complex vehicle routing and supply chain problems with quantum advantage.

---

[44] https://open-quantum-institute.cern

[45] https://www.gesda.global/open-quantum-institute-operations-kick-off-at-cern-with-continued-support-from-gesda-and-ubs/

[46] https://www.sipri.org/publications/2025/other-publications/military-and-security-dimensions-quantum-technologies-primer

[47] https://www.sipri.org/publications/2025/sipri-background-papers/introduction-military-quantum-technology-policymakers

[48] https://breakingdefense.com/2024/07/from-ukraine-to-taiwan-jamming-of-50-year-old-gps-is-a-defense-tech-nightmare/

[49] https://www.nato.int/en/news-and-events/articles/news/2024/01/17/nato-releases-first-ever-quantum-strategy

- **Materials design:** Accelerating the development of stronger, lighter aerospace, armour and lethal materials.
- **Weapons design:** Simulating complex physical phenomena that are impossible on classical computers.
- **Strategic simulation:** Modelling complex geopolitical and military scenarios with unprecedented fidelity.

The advent of operational quantum computing shifts the military-strategic balance. Militaries equipped with quantum computing capabilities will gain decisive advantages in logistics, procurement, and strategic planning. This advantage compounds over time.

### 5.3.Quantum-resistant cryptography: Defending against future threats

Post-quantum cryptography represents the most immediate quantum-related defence imperative. The transition from current cryptographic standards to quantum-resistant algorithms must be completed before quantum computers become operational. As already written earlier, most analysts agree that quantum computers will achieve sufficient capability by 2035, implying that PQC migration must accelerate immediately.

Defence-critical communications include:

- **Strategic command and control communications:** Ensuring the deployed forces' connections remain secure.
- **Intelligence dissemination:** Protecting classified assessments and operational intelligence.
- **NATO interoperability communications**: Ensuring secure, integrated communications.
- **Weapons systems:** Securing increasingly networked defence platforms.

Each category demands quantum-resistant protection. The challenge intensifies with legacy systems that may require replacement rather than upgrade.

## 6. The post-quantum cryptography imperative

### 6.1.The harvest now, decrypt later threat

Intelligence services worldwide recognise that some encrypted data captured today possesses significant strategic value once quantum computers become available. Diplomatic cables from 2024 could inform geopolitical positioning in 2035. Military plans encrypted today could guide adversary strategy tomorrow. Commercial secrets captured now could enable industrial espionage for decades.

### 6.2.The NIST standards and regulatory adoption

NIST's standards on post-quantum cryptographic algorithms provide the foundation for migration:

| Algorithm | Characteristics | Primary Application |
|---|---|---|
| ML-DSA | Lattice-based, optimized for efficiency | General-purpose digital signatures |
| ML-KEM | Lattice-based key encapsulation | Key establishment |
| FN-DSA | Compact lattice-based signatures | Space-constrained applications |

| SLH-DSA | Hash-based, conservative approach | Fallback option if lattice assumptions fail |
|---|---|---|
| HQC | Error-correcting codes approach | Key establishment (backup for ML-KEM) |

These algorithms share a critical property: their security derives from mathematical problems that are assumed to be intractable for both classical and quantum computers. This property—unlike current ageing RSA or ECC, which quantum computers will break—provides long-term security against quantum threats.

Regulatory adoption is accelerating. The EU's ENISA (European Agency for Cybersecurity) has issued guidelines for the transition to post-quantum cryptography.[50, 51] Germany's BSI[52], France's ANSSI[53], and the UK's GCHQ[54] are issuing (mandatory) adoption roadmaps.

## 6.3.Transition costs and timeline

Rigorous analysis is not widespread, and there are just a few serious analyses. BCG's analysis provides clarity on transition economics. Organisations operating an annual IT budget of US$1 billion can complete the PQC transition for approximately US$25 million if initiated immediately; delaying until 2035 may double the cost to US$50 million.[55]

Timeline considerations prove equally critical. Historical cryptographic transitions required 10-16 years from standard approval to full legacy system retirement. The DES-to-AES transition required 16 years. The MD-to-SHA family transition required 10 years. PQC transition cannot be compressed below 10 years—meaning organisations must begin immediately to complete it before 2035.

IoT device transitions present specific challenges. BCG estimates that automotive manufacturers face PQC transition costs of US$400-750 million due to the complexity of updating embedded systems across vehicle fleets. The manufacturing, utilities, and transportation sectors face US$10-20 million in transitions. These costs increase dramatically if companies delay migration.

The U.S. General Services Administration published a useful *PQC Buyer's Guide*[56] in June 2025.

The NCSC guidance, which outlines a three-phase timeline for organisations to transition to quantum-resistant encryption methods by 2035, is also a valuable resource.[57]

# 7. Strategic recommendations for quantum-ready defence and policy institutions

## 7.1.For defence and government leaders

**1. Integrate quantum into strategic planning**

[50] https://digital-strategy.ec.europa.eu/en/library/coordinated-implementation-roadmap-transition-post-quantum-cryptography

[51] https://cyber.gouv.fr/sites/default/files/document/follow_up_position_paper_on_post_quantum_cryptography.pdf

[52] https://www.bsi.bund.de/EN/Themen/Unternehmen-und-Organisationen/Informationen-und-Empfehlungen/Quantentechnologien-und-Post-Quanten-Kryptografie/quantentechnologien-und-post-quanten-kryptografie_node.html

[53] https://cyber.gouv.fr/actualites/cryptographie-post-quantique-les-travaux-de-lanssi

[54] https://www.ncsc.gov.uk/news/pqc-migration-roadmap-unveiled

[55] https://www.bcg.com/publications/2025/how-quantum-computing-will-upend-cybersecurity

[56] https://buy.gsa.gov/api/system/files/documents/final-508c-pqc_buyer-s_guide_2025.pdf

[57] https://www.ncsc.gov.uk/news/pqc-migration-roadmap-unveiled

Treat quantum not as a technical curiosity but as a strategic domain equivalent to space or cyber. Allocate planning resources to the quantum strategy as you would to other critical domains. Establish quantum-specific career tracks and ensure continuous access to quantum expertise within strategic planning organisations.

**2. Balance post-quantum cryptography migration and quantum key distribution deployment**
Classify all government communications into criticality tiers. Any delay increases the vulnerability window.

**3. Ensure NATO interoperability**
Work within NATO frameworks to establish standardised quantum-resistant cryptographic approaches, ensuring allied communications remain secure and interoperable. Current NATO interoperability standards assume classical cryptography; quantum-ready interoperability requires deliberate standards development and implementation.

**4. Establish quantum sensing integration**
Begin integrating quantum-capable sensing systems into military operations, starting with pilot programs. Establish doctrine for quantum-enabled sensing employment. Recognise that potential adversaries developing quantum sensing capabilities pose strategic challenges that require matched development.

## 7.2.For industry and standards leaders

**1. Develop crypto agility**
Shift cryptographic architectures from rigid, embedded designs to modular, agile systems enabling rapid algorithm substitution as quantum threats evolve. NIST's March 2025 guidance on crypto agility[58] provides a blueprint. The EC's June 2025 "A Coordinated Implementation Roadmap for the Transition to Post-Quantum Cryptography"[59] offers excellent advice. Implement automated mechanisms enabling rapid cryptographic standard updates without system redesign.

**2. Proactively communicate PQC timelines**
Organisations managing critical infrastructure, supply chains, and security-sensitive services should communicate PQC migration timelines to customers and partners. Transparency regarding transition plans builds trust and enables coordinated industry transitions.

**3. Participate in standards development**
Engage actively in ETSI, ISO, IEEE and other standards bodies developing quantum standards. Standards decisions implicitly determine competitive positioning—organisations that shape standards gain a competitive advantage. It is recommended to prefer global standards organisations that are immune to geopolitics, such as ETSI.

## 7.3.For research and academic institutions

**1. Expand quantum education and workforce development**
Quantum expertise remains scarce. Academic institutions should expand programs in quantum physics, quantum engineering, and quantum computer science to build a workforce capable of developing and deploying quantum technologies. A few European initiatives, such as QUARMEN[60] and QuanTEEM[61], are excellent examples of joining forces to deliver high-quality education.

---

[58] https://csrc.nist.gov/pubs/cswp/39/considerations-for-achieving-cryptographic-agility/2pd

[59] https://digital-strategy.ec.europa.eu/en/library/coordinated-implementation-roadmap-transition-post-quantum-cryptography

[60] https://www.master-quarmen.eu

[61] https://www.quanteem.eu

**2. Maintain open publication in basic quantum science**
While applied quantum technologies warrant appropriate security classification, fundamental quantum science benefits from open publication and international collaboration. Maintain the scientific openness that has historically accelerated scientific progress while implementing appropriate security measures around dual-use applications. Applying the '*As open as possible, as closed as needed*' principle might be wise.

## 8. Conclusion: The quantum-ready future

Quantum technologies represent a singular inflexion point in technological competition and geopolitical strategy. Unlike previous technological revolutions that extended existing capabilities, quantum introduces fundamentally new physical principles enabling capabilities previously impossible and threatening existing security measures.

The strategic imperative is clear: organisations that act decisively to integrate quantum capabilities into strategy, accelerate the migration to post-quantum cryptography, and establish quantum-ready defence and diplomatic frameworks will secure strategic advantage.

Quantum is, by design, a long-term play because the most disruptive capabilities—particularly in quantum computing—sit at the intersection of complicated physics, complex engineering, and fragile supply chains, all of which evolve on decade-long horizons rather than budget cycles. As explained, quantum computers that materially outperform classical systems at scale will require fault-tolerant architectures, industrialised fabrication, and mature software stacks; leading roadmaps still point to the second half of the 2030s for broadly applicable, general-purpose systems, even as niche advantages emerge earlier in chemistry, optimisation, and codebreaking.

For governments and armed forces, this means quantum cannot be treated as a tactical "pilot" topic, but as an infrastructure-level bet that must be stewarded over multiple administrations and force-planning cycles. A credible strategy starts with a long-term mission view: identifying the specific problem classes—submarine detection, logistics optimisation, codebreaking, autonomous systems—where quantum advantage would reshape deterrence and operational concepts, and then funding stable pipelines that connect basic research, dual-use industry, and defence testbeds over 15–20 years. It also requires acting now on "slow variables" such as talent, secure supply chains, export control positions, and post-quantum cryptography migration, where today's decisions determine whether, in the 2030s, a state is a quantum taker or a quantum shaper.

Quantum is not just another emerging technology to monitor; it is a structural capability race, and the winners will be those governments and defence establishments that treat it as a sustained portfolio of bets, with clear mission outcomes, institutional ownership, and resilience to political and budgetary noise.

For defence leaders, the quantum challenge manifests in four specific imperatives:

1. **Accelerate investment** in quantum capabilities across sensing, communication, and computing while maintaining technological pluralism, enabling multiple approaches to quantum innovation. A careful '*Technology Readiness Levels*' approach is recommended.

2. **Establish quantum diplomacy frameworks** that engage allies and neutral parties to shape standards and interoperability approaches, ensuring that allied systems remain compatible despite strategic competition. These frameworks must balance specific supply chain needs (such as raw materials like helium-3 and Silicon-28, critical technologies like FPGAs, and the consolidation of manufacturing capabilities for specific quantum components).

3. **Prioritise migrating to post-quantum cryptography to** protect classified communications, defence systems, and strategic infrastructure against "harvest now, decrypt later" threats. Europe at large, NATO, the European Union, the United Kingdom and the United States must work together in a coordinated manner and find the right balance between sovereignty and dependence. What is at stake is not just adopting global standards but also how they are implemented.

4. **Integrate quantum capabilities** into defence strategy and military operations, recognising that quantum-enabled sensing, computing, and communication will reshape military advantage.

We do not operate in an ideal world. It is precisely because of this imperfect reality that prioritising a coordinated and collaborative transition—balancing post-quantum cryptography migration with quantum key distribution deployment—is imperative to securing global critical infrastructure, including satellites. This must be a foremost concern for policymakers and defence leaders committed to safeguarding the future.

The quantum era has commenced. The race is well underway. Organisations that recognise quantum's strategic significance and act proactively and decisively will thrive in the quantum-enabled future. Those that delay will discover, to their cost, that quantum advantages compound over time—and that falling behind proves extraordinarily difficult.